\documentclass[twocolumn,aps,prl,showpacs,amsmath,amssymb,superscriptaddress]{revtex4-1}

\usepackage[usenames]{color}
\usepackage{hyperref}
\usepackage{graphicx}

\def\<{\langle}
\def\>{\rangle}

\begin{document}

\title{Dynamical formation of the unitary Bose gas}

\author{V. E. Colussi}
\affiliation{Eindhoven University of Technology, PO Box 513, 5600 MB Eindhoven, The Netherlands}
\author{S. Musolino}
\affiliation{Eindhoven University of Technology, PO Box 513, 5600 MB Eindhoven, The Netherlands}
\author{S. J. J. M. F. Kokkelmans}
\affiliation{Eindhoven University of Technology, PO Box 513, 5600 MB Eindhoven, The Netherlands}

\begin{abstract}  
We study the structure of a Bose-condensed gas after quenching interactions to unitarity.  Using the method of cumulants, we decompose the evolving gas in terms of clusters.  Within the quantum depletion we observe the emergence of two-body clusters bound purely by many-body effects, scaling continuously with the atomic density.  As the unitary Bose gas forms, three-body Efimov clusters are first localized and then sequentially absorbed into the embedded atom-molecule scattering continuum of the surrounding depletion.  These results motivate future experimental probes of a quenched Bose-condensate during evolution at unitarity. 


\end{abstract}

\maketitle 
{\it Introduction.}---Precision control of external magnetic fields allows ultracold Bose gas experiments to tune interactions, characterized by the s-wave scattering length $a$.  Via Feshbach resonances \cite{RevModPhys.82.1225}, experiments have accessed the degenerate unitary regime $n|a|^3\to\infty$ with atomic density $n$, beating per-particle losses scaling as $\dot{n}/n\sim n^2a^4$ by diabatically quenching the scattering length to resonance ($|a|\to\infty$) \cite{makotyn2014universal,klauss2017observation,eigen2017universal}.  The insensitivity of unitary quantum gases to diverging microscopic length scales extends their properties to seemingly unrelated strongly-correlated physical systems, such as the inner crust of neutron stars and the quark-gluon plasma \cite{0034-4885-72-12-126001}.  This predictive power is due to the intrinsic scale invariance of these unitary systems  \cite{chin2017ultracold}.  Strong experimental evidence for two-component unitary Fermi gases \cite{Hara2179,Cao58} supports a universal thermodynamics based solely on continuous power laws of the atomic density derived ``Fermi'' scales $k_\mathrm{n}=(6\pi^2n)^{1/3}$, $E_\mathrm{n}=\hbar^2k_\mathrm{n}^2/2m$, and time $t_\mathrm{n}=\hbar/E_\mathrm{n}$ where $m$ is the atomic mass \cite{PhysRevLett.92.090402}.  The scaling behavior of the unitary Bose gas is complicated by the finite-size and discrete-scaling properties of three-body bound Efimov states \cite{efimov1971weakly,efimov1979low}, introducing a complex scaling dimension \cite{chin2017ultracold}.  A full characterization of the quasi-equilibrium state of the unitary degenerate Bose gas observed experimentally \cite{makotyn2014universal,eigen2017universal} remains an open question.  
 
 These difficulties are symptoms of an undeveloped picture of few-body physics in the evolving many-body background and their manifestations in system properties on Fermi timescales.  Recently, the problem of merging the Efimov effect and a many-body background has received attention in the related context of impurities immersed in static bosonic \cite{doi:10.7566/JPSJ.87.043002,PhysRevLett.119.013401,PhysRevX.8.011024,0295-5075-101-6-60009} or fermionic  \cite{PhysRevA.94.051604,1367-2630-16-2-023026,PhysRevLett.106.166404,PhysRevLett.106.145301,PhysRevA.93.053611,PhysRevLett.114.115302} media.  However, the dynamical nature of quench experiments poses an additional theoretical challenge.  Initially, the quench disturbs short-range physics in the gas, inducing ballistic correlation waves \cite{PhysRevA.94.023604} and sequential clustering \cite{PhysRevLett.120.100401,josearxiv}.  Recently measured per-particle-loss rates for quenched unitary Bose gases scaling continuously over a range of atomic densities suggest that Efimov physics plays only a minor role for this observable \cite{klauss2017observation,eigen2017universal}.  However, over a wider range of atomic densities, preliminary loss-rate measurements \cite{klaussthesis} and theoretical results \cite{josearxiv} indicate a log-periodic oscillation of the loss-rate with a density period set by the Efimov spacing $e^{3\pi/s_0}\approx 22.7^3$ where $s_0\approx1.00624$ is a universal constant for three identical bosons \cite{efimov1971weakly}.  These results parallel oscillatory loss-rate predictions in the nondegenerate regime \cite{PhysRevX.6.021025}.
 
In this Letter, we explore the composition of a Bose-condensate quenched to unitarity.  Our model applies to broad, entrance-channel dominated Feshbach resonances that are well-approximated by short-range single-channel interactions \cite{RevModPhys.82.1225}.  This system has been realized experimentally in Refs.~\cite{makotyn2014universal,klauss2017observation} using $^{85}$Rb and in Ref.~\cite{eigen2017universal} using $^{39}$K.  Using the method of cumulants, we derive two- and three-body Schr\"odinger equations including density effects.  These yield the evolving spectrum of bound two- and three-body clusters.  We map out the dynamical and density scaling properties of the bound cluster spectrum and comment on manifestations in system properties. 

{\it Cumulant equations.}---
Our quantitative many-body theory of the Bose-condensed gas quenched to unitarity is built from the cumulant expansion, which classifies correlated particle clusters within an interacting many-body system \cite{kohler2002microscopic,kira2011semiconductor}.  The second-order cumulant expansion yields the Hartree-Fock-Bogoliubov equations (HFB) \cite{blaizot1986quantum}.  These equations may be systematically extended to higher order, yielding few-particle cluster kinetics that can be used to explore strongly-interacting few-body physics like the Efimov effect.  In terms of the bosonic annihilation and creation operators, $\hat{a}_{\bf k}$ and $\hat{a}_{\bf k}^\dagger$ respectively, for a particle of momentum $\hbar {\bf k}$, cumulants are defined from normal-ordered expectation values
\begin{align}
\left\langle \prod_{i=1}^l\hat{a}^\dagger_{{\bf k}_i} \prod_{j=1}^m\hat{a}_{{\bf q}_j}\right\rangle_c&\equiv(-1)^m \prod_{i=1}^l\frac{\partial}{\partial x_i}\prod_{j=1}^m\frac{\partial}{\partial y_j^*}&\nonumber\\
\times &\ln\left.\left\langle e^{\sum_{i=1}^l x_i \hat{a}^\dagger_{{\bf k}_i}} e^{-\sum_{j=1}^m y_j^* \hat{a}_{{\bf q}_j}}\right\rangle\right|_{
{\bf x},{\bf y}=0},&\nonumber\\
\end{align}
in terms of complex-valued $x_i$ and $y_j$.  For uniform systems, the set of relevant cumulants in the above equation are restricted such that $\sum_{i=1}^l{\bf k_i}=\sum_{j=1}^m{\bf q_j}$.  To model the condensate and excitations, we make the Bogoliubov approximation \cite{blaizot1986quantum}, decomposing operators as $\hat{a}_{\bf k}=\psi_{\bf k}+\delta\hat{a}_{\bf k}$ in terms of coherent state amplitude $\langle \hat{a}_{\bf k}\rangle=\psi_0\delta_{{\bf k},0}$ and fluctuations $\langle\delta\hat{a}_{{\bf k}\neq 0}\rangle=0$.  This is justified provided excited modes are not macroscopically occupied.  Isolating the condensate in the first-order cumulant $\psi_0$, we truncate the cumulant expansion at second-order, which describes genuine two-excitation correlations.  This includes also the one-body $\rho_{\bf k}\equiv\langle \hat{a}^\dagger_{\bf k}\hat{a}_{\bf k}\rangle_c$ and pairing $\kappa_{\bf k}\equiv\langle \hat{a}_{\bf k}\hat{a}_{\bf -k}\rangle_c$ density matrices for excitations.  

\begin{figure}[t!]
\includegraphics[width=8.6cm]{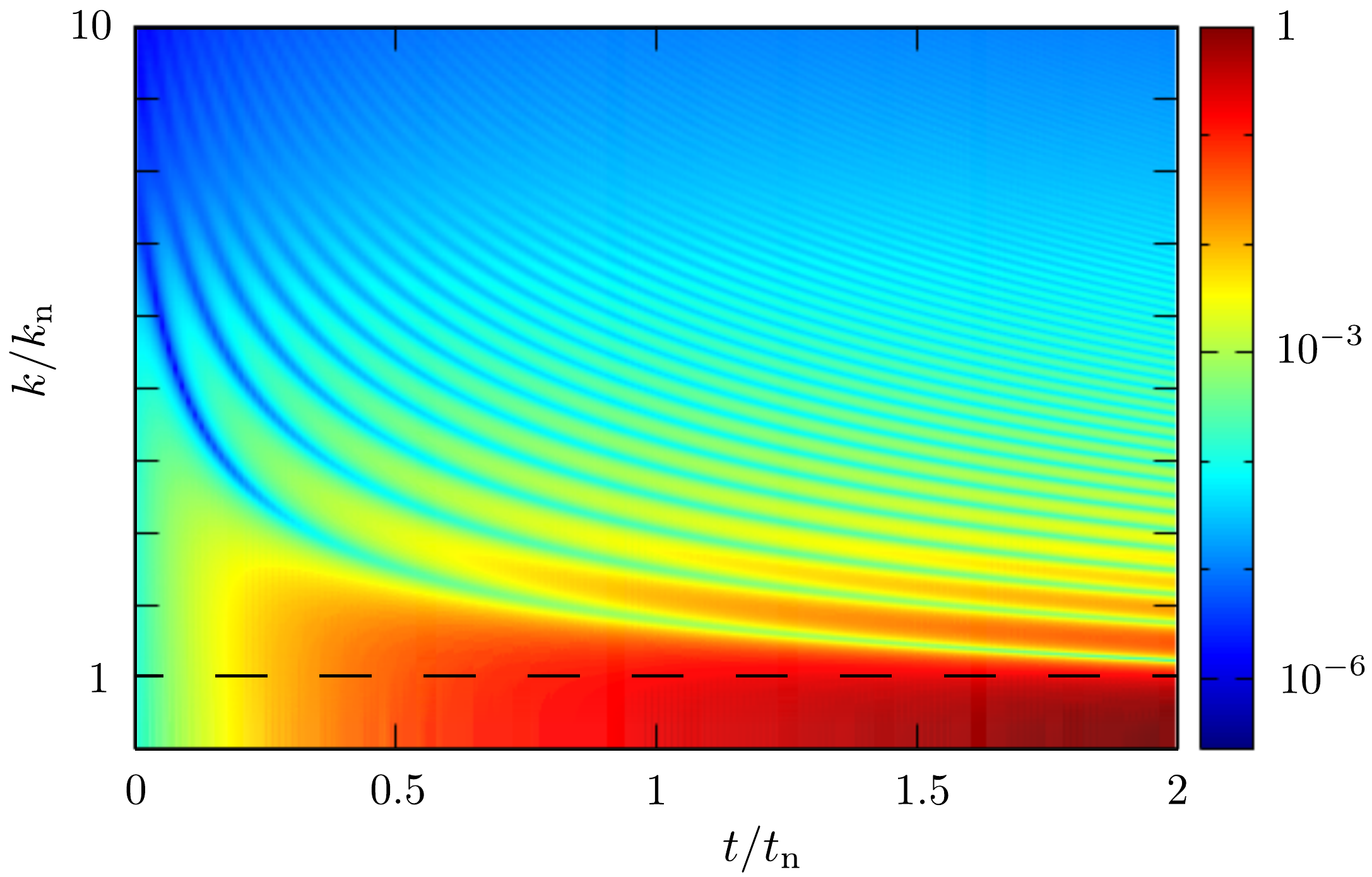}
\caption{\label{fig:gn} Density plot of the universal excitation density $\rho_{\bf k}$ evolving at unitarity after a $5$ $\mu$s quench.  The ``rippling'' effect is due to ballistic correlation waves studied in Ref.~\cite{PhysRevA.94.023604}.  The dashed line indicates the scale of the Fermi wavenumber $k_\mathrm{n}$ where excitation buildup is most pronounced,  and $\rho_{\bf k}>1$ for $t\gtrsim2t_\mathrm{n}$.}
\end{figure}
We utilize a single-channel many-body Hamiltonian applicable in the vicinity of a broad Feshbach resonance
\begin{align}\label{eq:mbham} 
\hat{H}&=\sum_{\bf k}\frac{\hbar^2 k^2}{2m}\hat{a}^\dagger_{\bf k}\hat{a}_{\bf k}&\nonumber\\
&+\frac{g}{2}\sum_{{\bf p},{\bf p'},{\bf q}}\zeta({\bf p}-{\bf p'}+2{\bf q})\zeta^*({\bf p}-{\bf p'}) \hat{a}^\dagger_{{\bf p}+{\bf q}}\hat{a}^\dagger_{{\bf p'}-{\bf q}}\hat{a}_{\bf p} \hat{a}_{\bf p'}.&\nonumber\\
\end{align}  
At energies close to a two-body bound-state, the two-body T-matrix becomes separable \cite{faddeev2013quantum}, and we use a non-local separable pairwise potential, $\hat{V}=g|\zeta\rangle\langle\zeta|$.  We employ a step-function form factor $\zeta({\bf k})=\Theta(\Lambda-|{\bf k}|/2)$, which has been previously used to study Efimov states in vacuum (cf. Ref.~\cite{0034-4885-80-5-056001}).  The s-wave interaction strength $g$ is calibrated to reproduce the zero-energy limit of the two-body T-matrix $g=U_0\Gamma$ where $U_0=4\pi\hbar^2a/m$ and $\Gamma=(1-2a\Lambda/\pi)^{-1}$.  In the $\Lambda\to\infty$ limit, $\hat{V}$ is equivalent to a renormalized contact potential, however we do not take $\Lambda$ arbitrarily large.  In the spirit of Refs.~\cite{PhysRevLett.89.180401,PhysRevA.76.012720,PhysRevA.72.022714}, $\Lambda$ is instead calibrated to reproduce finite-range corrections to the Feshbach molecule binding energy $-\hbar^2/m(a-\bar{a})^2$ away from unitarity where $\bar{a}=0.955r_\mathrm{vdW}$ is the mean-scattering length depending on the van der Waals length $r_\mathrm{vdW}$ for a particular atomic species \cite{PhysRevA.48.546,RevModPhys.82.1225}---see Supplemental Material \cite{SM} for $^{39}$K and $^{85}$Rb calibration.  This yields $\Lambda=2/\pi\bar{a}$, introducing finite-range effects into our many-body model, removing the need for an additional three-body parameter, and avoiding the unphysical Thomas collapse \cite{PhysRev.47.903} in our calculation of Efimov clusters discussed below.    

From Eq.~(\ref{eq:mbham}), we use the Heisenberg equation of motion $i\hbar\dot{\hat{\mathcal{O}}}=[\hat{\mathcal{O}},\hat{H}]$ and obtain the HFB equations for the dynamics of the first and second-order cumulants 
\begin{align}
i\hbar\dot{\psi}_0&=g\left(|\zeta(0)|^2|\psi_0|^2+2\sum_{{\bf k\neq}0}|\zeta({\bf k})|^2\rho_{\bf k}\right)\psi_0&\nonumber\\
&+g\psi_0^* \sum_{{\bf k}\neq0}\zeta(0)\zeta^*(2{\bf k})\kappa_{\bf k},&\label{eq:gpe}\\
i\hbar\dot{\kappa}_{\bf k}&=2h({\bf k})\kappa_{\bf k}+(1+2\rho_{\bf k})\Delta({\bf k}),&\label{eq:kappadot}\\
\hbar\dot{\rho}_{\bf k}&=2\ \text{Im} \left(\Delta({\bf k})\kappa_{\bf k}^*\right),&\label{eq:rhodot}
\end{align}
where
\begin{align}
h({\bf k})&=\frac{\hbar^2k^2}{2m}+2g|\zeta({\bf k})|^2|\psi_0|^2+2g\sum_{{\bf k'}\neq 0}|\zeta({\bf k-k'})|^2\rho_{\bf k'},& \label{eq:hfham}\\
 \Delta({\bf k})&=g\zeta(2{\bf k})\left(\zeta^*(0)\psi_0^2+\sum_{{\bf k'}\neq0}\zeta^*(2{\bf k'})\kappa_{\bf k'}\right),&\label{eq:pairing}
 \end{align}
are the Hartree-Fock Hamiltonian and pairing field, respectively \cite{blaizot1986quantum}.  We mimic the initial quench sequence of Refs.~\cite{makotyn2014universal,klauss2017observation,eigen2017universal} and ramp a pure Bose-condensate onto resonance over the course of $5$ $\mu$s and then evolve the system at unitarity.  The HFB theory, Eqs.~(\ref{eq:gpe})--(\ref{eq:rhodot}), describes the quantum depletion of a Bose-condensate via the generation of correlated excitation pairs studied in Ref.~\cite{PhysRevA.89.021601}.  The universal evolution of the excitation density $\rho_{\bf k}$ is shown in Fig.~\ref{fig:gn}, where a decaying $k^{-4}$ leading-order tail develops at high momentum proportional to the Tan contact \cite{TAN20082952,TAN20082971,TAN20082987}.  This is due to the growth of two-body correlations at short-distances $r\ll n^{-1/3}$ \cite{PhysRevA.89.021601}.   On the Fermi timescale, a macroscopic buildup of excitations occurs on the scale of $k_\mathrm{n}$, indicated by the dashed line in Fig.~\ref{fig:gn}, eventually violating the assumptions underlying our model as $\rho_{\bf k}$ exceeds unity.  We find that this breakdown occurs universally after evolving a time $t\approx 2t_\mathrm{n}$ at unitarity.  

{\it Embedded few-body Schr\"odinger equations.}---
Equations~(\ref{eq:gpe})--(\ref{eq:rhodot}) describe the evolving many-body background up to second-order correlations.  Using this description, we investigate the bound two- and three-body clusters formed within the depletion and introduce to the set of cumulant equations the triplet $\tau^{0,3}_{\bf k,k'}=\langle \hat{a}_{\bf -k-k'}\hat{a}_{\bf k}\hat{a}_{\bf k'}\rangle_c$, where the superscript notation indicates the number of creation and annihilation operators, respectively.  Unlike the embedded impurity problem, bound clusters in the unitary Bose-condensed gas are indistinguishable from the background and are therefore subject to Bose-stimulation.  The dynamics of $\kappa_{\bf k}$ and $\tau^{0,3}_{\bf k,k'}$, which include two- and three-body scattering in medium, generally occur on timescales shorter than the density dynamics \cite{KIRA2015185,kira2011semiconductor}.  Treating density effects as quasi-stationary, the principle portion of the cumulant equations for $\kappa_{\bf k}$ and $\tau^{0,3}_{\bf k,k'}$ defines eigenvalue equations 
\begin{widetext}
\begin{align}  
E^{(\nu)}_\mathrm{2B}\phi_\nu({\bf k})&=2h({\bf k})\phi_\nu({\bf k})+(1+2\rho_{\bf k})\sum_{{\bf k'}\neq 0}g\zeta(2{\bf k})\zeta^*(2{\bf k'})\phi_\nu({\bf k'}),&\label{eq:wannier}\\
E^{(\nu)}_\mathrm{3B}\Psi_{\nu}({\bf k},{\bf k'})&=(1+\hat{P}_++\hat{P}_-)\left(h({\bf k})\Psi_{\nu}({\bf k},{\bf k'})+(1+\rho_{\bf k'}+\rho_{{\bf k}+{\bf k'}})\sum_{{\bf k''}\neq 0}g\zeta(2{\bf k'}+{\bf k})\zeta^*(2{\bf k''}+{\bf k})\Psi_\nu({\bf k},{\bf k''})\right),&\label{eq:genthree}
\end{align}
\end{widetext}
where we have ignored inhomogeneities that describe scattering amongst clusters (see Ref.~\cite{SM}).  Iterative solution of Eqs.~(\ref{eq:wannier})--(\ref{eq:genthree}) yields two- and three-body cluster eigenenergies $E^{(\nu)}_\mathrm{2B}$ and $E^{(\nu)}_\mathrm{3B}$ and right-handed wave functions  $\phi_\nu({\bf k})$ and $\Psi_{\nu}({\bf k},{\bf k'})$ evolving on the timescale of the many-body background \cite{handedness}.  This treatment is formally similar to the derivation of the hyperbolic Wannier equation \cite{KIRA2015185}, and both equations are bosonic analogues of the Wannier equation \cite{PhysRev.52.191,kira2011semiconductor} describing bound electron-hole pairs in semiconductors.  The operators $\hat{P}_-$, $\hat{P}_+$ indicate cyclic and anti-cyclic permutations, respectively.  In the zero-density limit, Eqs.~(\ref{eq:wannier})--(\ref{eq:genthree}) reduce to the two- and three-particle Schr\"odinger equation, respectively, and therefore describe {\it embedded} extensions.  

In the regime $\Lambda\gg\xi^{-1}$, where $\xi^2=\hbar^2/2m|g|n$ is the condensate healing length \cite{doi:10.1063/1.1703944},  we find that $h({\bf k})\approx \hbar^2 k^2/m+2gn$, and the structure of Eqs.~(\ref{eq:wannier})--(\ref{eq:genthree}) simplifies.  In our model at unitarity this limit is equivalent to the diluteness criterion $nr_\mathrm{vdW}^3\ll1$, which is well satisfied by all unitary degenerate Bose gas experiments to date ($nr_\mathrm{vdW}^3<10^{-5}$) \cite{makotyn2014universal,klauss2017observation,eigen2017universal}.  Consequently, we report cluster binding energies $\tilde{E}^{(\nu)}_\mathrm{2B}\equiv E^{(\nu)}_\mathrm{2B}-4gn$ and $\tilde{E}^{(\nu)}_\mathrm{3B}\equiv E^{(\nu)}_\mathrm{3B}-6gn$ relative to the embedded two- and three-body continuum thresholds, with $g=-\pi^3\hbar^2\bar{a}/m$ in the unitary regime.  Additionally, we define a nonsymmetric effective pairwise interaction $\hat{V}_\mathrm{eff}\equiv\hat{B}\hat{V}$ where $\langle {\bf k,k'}|\hat{B}=(1+\rho_{\bf k}+\rho_{\bf k'})\langle {\bf k,k'}|$ Bose-enhances collisions occurring in medium.  On the Fermi timescale, the operator $\hat{B}$ enhances pairwise interactions disproportionately at the scale of the inter-particle spacing, as shown in Fig.~\ref{fig:gn}.  This effect was first studied in Ref.~\cite{KIRA2015185} for a Bose-condensed gas of $^{85}$Rb quenched to unitarity at density $nr_\mathrm{vdW}^3=2\times10^{-7}$ and evolution time $t\sim800$ $\mu$s, observing a $528$ Hz ($\approx0.3E_\mathrm{n}$) blueshift in the binding energy of the resonant two-body bound state \cite{phen}.  In this Letter, we present a systematic study of the evolution of two- and three-body bound clusters in the unitary regime over a range of densities. 

 \begin{figure}[t!]
\includegraphics[width=8.6cm]{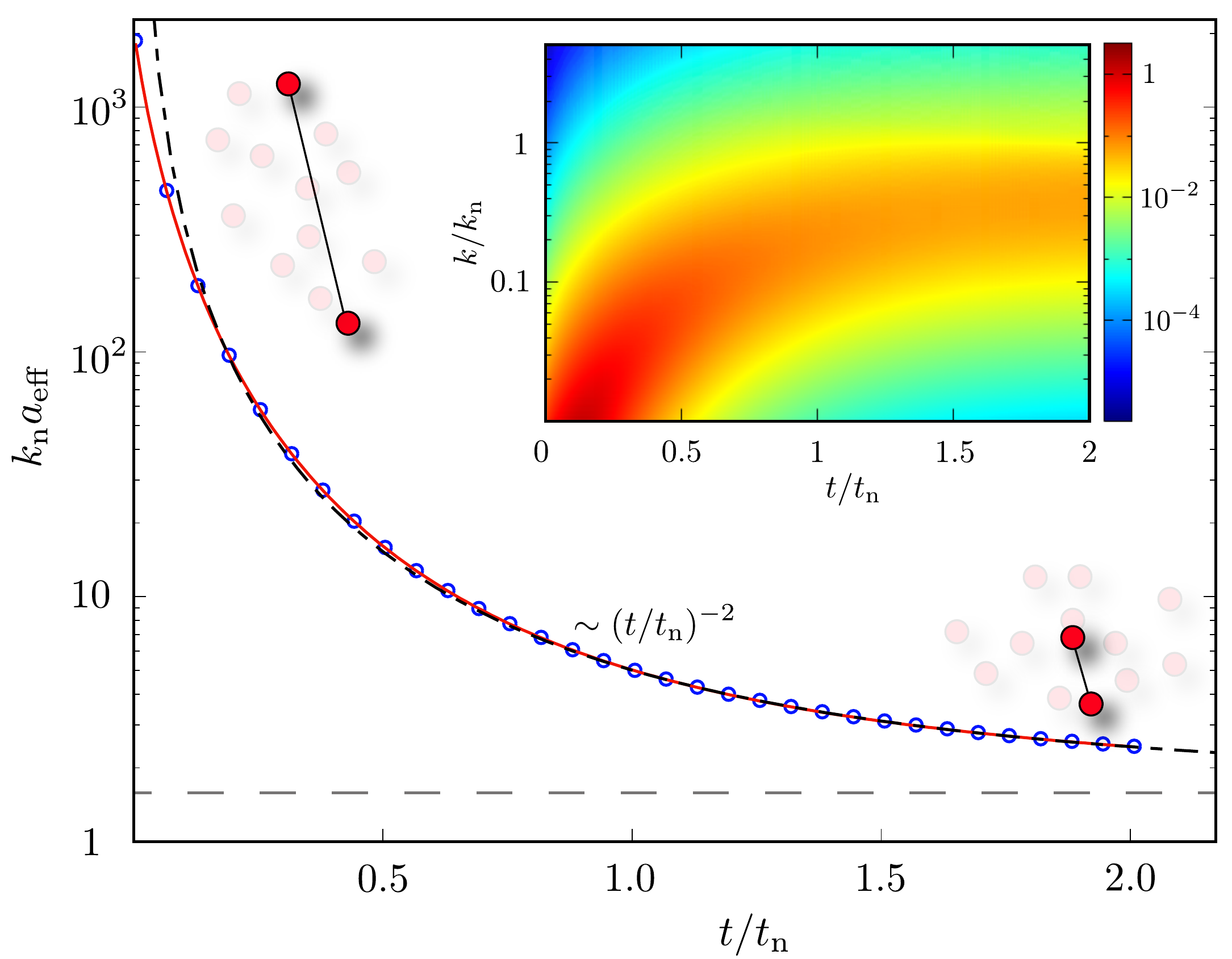}
\caption{\label{fig:aeff}  Evolution of $a_\mathrm{eff}$ for two densities within the range of experimental interest, $nr_\mathrm{vdW}^3=10^{-7}$ (solid red curve) and $10^{-9}$ (blue circles).  The fitted universal result in Eq.~(\ref{eq:universaldimer}) corresponds to the dash-dotted line with asymptotic estimate $a_\mathrm{eff}=0.41 n^{-1/3}$ indicated by the dashed line.  The inset shows a density plot of the universal dynamics of the two-body bound cluster probability density $k^2|\phi_\mathrm{\mathrm{D}}({\bf k})|^2/(1+2\rho_{\bf k})$ in arbitrary units \cite{handedness}.}
\end{figure}
{\it Two-body bound clusters.}---To study bound two-body clusters, we reformulate the embedded two-body Schr\"odinger equation, Eq.~(\ref{eq:wannier}), as a Lippman-Schwinger equation for the embedded two-body T-operator $\hat{\mathcal{T}}_\mathrm{2B}(z)=\hat{B}\hat{V}+\hat{B}\hat{V}\hat{G}_\mathrm{2B}^{(0)}(z)\hat{\mathcal{T}}_\mathrm{2B}(z)$, where $\hat{G}_\mathrm{2B}^{(0)}(z)\equiv(z-2\hat{t})^{-1}$ is the two-body free Green's operator with kinetic energy operator $\hat{t}|{\bf k}\rangle=\hbar^2k^2/2m|{\bf k}\rangle$ and energy $z$ relative to the embedded two-body continuum threshold (see Ref.~\cite{SM}).  Our $\hat{\mathcal{T}}_{2B}(z)$ is related to the ``many-body T-operator'' $\hat{B}\hat{T}_\mathrm{MB}(z)=\hat{\mathcal{T}}_\mathrm{2B}(z)$ introduced in Ref.~\cite{stoof1996theory}, which predicts weakly bound pairs at unitarity in the finite temperature phase diagram of the strongly-interacting Bose gas \cite{PhysRevA.79.063609}.  For separable potentials, we obtain the closed expression  
\begin{equation}\label{eq:embeddedt} 
\hat{\mathcal{T}}_\mathrm{2B}(z)=\hat{B}\frac{g|\zeta\rangle\langle\zeta|}{1-g\langle \zeta|\hat{G}_\mathrm{2B}^{(0)}(z)\hat{B}| \zeta\rangle}.
\end{equation}
The position of the simple pole in Eq.~(\ref{eq:embeddedt}) corresponds to the dimer binding energy of a two-body cluster $z=\tilde{E}^{(\mathrm{D})}_\mathrm{2B}$, with wave function $|\phi_\mathrm{D}\rangle\propto \hat{G}_\mathrm{2B}^{(0)}(\tilde{E}^{(\mathrm{D})}_\mathrm{2B})\hat{B}|\zeta\rangle$.   

To parametrize the binding energy and size of the two-body bound cluster, we define an effective two-body scattering length $-\hbar^2/ma_\mathrm{eff}^2\equiv \tilde{E}^{(\mathrm{D})}_\mathrm{2B}$.  Over a range of densities and times shown in Fig.~\ref{fig:aeff}, we find that $a_\mathrm{eff}$ scales continuously solely with the density quantified by the dynamical scaling power law 
\begin{equation}\label{eq:universaldimer}
\tilde{E}_\mathrm{2B}^{(\mathrm{D})}=-E_\mathrm{n}\left(1.12+2.43\left(\frac{t_\mathrm{n}}{t}\right)^2\right)^{-2}.
\end{equation}
This fitted equation matches the universal binding energy $\tilde{E}_\mathrm{2B}^{(D)}\approx -0.3E_\mathrm{n}$ found at the latest time considered in our model $t\approx 2t_\mathrm{n}$ and predicts the universal asymptotic binding energy $\tilde{E}_\mathrm{2B}^{(\mathrm{D})}\approx -0.8E_\mathrm{n}$.  Due to the minimal amount of quantum depletion during the quench, the two-body bound cluster is initially nearly-resonant $k_\mathrm{n}a_\mathrm{eff}\sim 10^3$ with the embedded two-body scattering threshold as shown in Fig.~\ref{fig:aeff}.  Quantum depletion on the Fermi timescale enhances pairwise interactions at the scale of $k_\mathrm{n}$ shown in Fig.~\ref{fig:gn}, and $\hat{V}_\mathrm{eff}$ supports a universal two-body cluster bound entirely by many-body effects.  Consequently,  the extended two-body bound cluster shrinks to the asymptotic prediction $a_\mathrm{eff}=0.41n^{-1/3}$ of Eq.~(\ref{eq:universaldimer}).  The dynamic localization of the universal bound two-body cluster towards the scale of the interparticle spacing is shown in the inset of Fig.~\ref{fig:aeff}.

 \begin{figure}[t!]
\includegraphics[width=8.6cm]{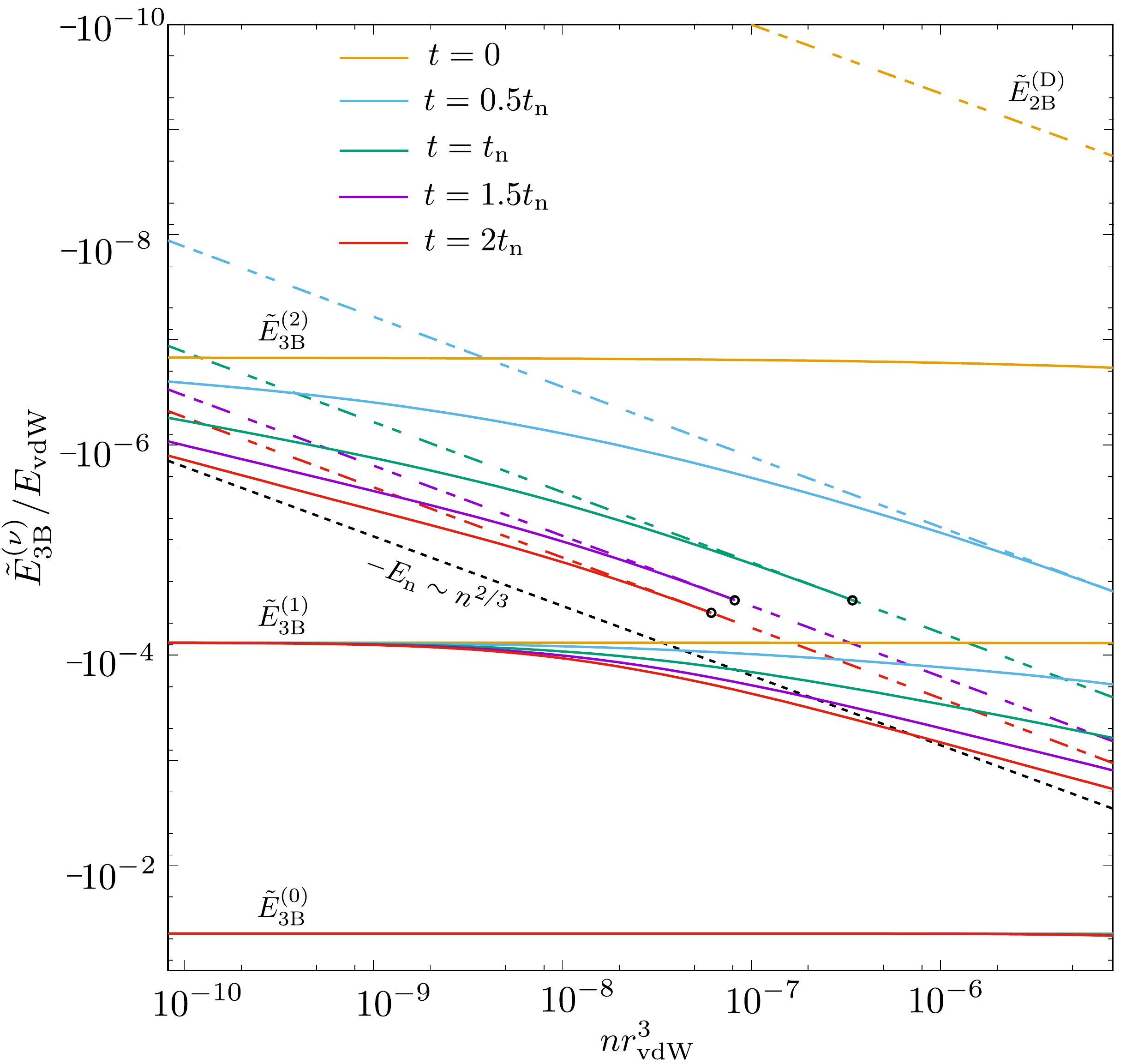}
\caption{Efimov cluster (solid) and universal two-body cluster (dot-dashed) binding energies over a range of evolution times at unitarity and densities of experimental interest.  The circled data points indicate the absorption of an Efimov cluster into the embedded atom-molecule threshold.   The log-log scale reveals the scaling behavior of the energies with the gas density and the Fermi energy (dashed line). \label{fig:energy_scaling}}
\end{figure}
%
{\it Three-body bound clusters.}---In vacuum it is well-known that the shallow two-body bound state for $a>0$ is associated with a finite set of Efimov states, merging sequentially with the atom-molecule threshold as $a$ is decreased from unitarity \cite{BRAATEN2007120}.  Analogously, the dynamical formation of the universal bound two-body cluster and coincident decrease of $a_\mathrm{eff}$ must also have consequences for the spectrum of Efimov clusters.  

To study these effects, we decompose the three-body wave function into Faddeev components $|\Psi_\nu\rangle=(1+\hat{P}_++\hat{P}_-)|\Psi_\nu^{(1)}\rangle$, obeying the bound-state Faddeev equation $|\Psi_\nu^{(1)}\rangle=\hat{G}_\mathrm{3B}^{(0)}(z)\hat{\mathcal{T}}_{23}(z)(\hat{P}_++\hat{P}_-)|\Psi_\nu^{(1)}\rangle$ \cite{faddeev2013quantum} where $\hat{\mathcal{T}}_{23}(z)=\hat{B}_1\hat{V}_1+\hat{B}_1\hat{V}_1\hat{G}_\mathrm{3B}^{(0)}(z)\hat{\mathcal{T}}_{23}(z)$, and the energy $z$ is defined relative to the embedded three-body continuum threshold.  Here we have used the spectator notation to indicate pairwise interaction between atoms $2$ and $3$ and defined the three-body free Green's operator, $\hat{G}_\mathrm{3B}^{(0)}(z)\equiv(z-\sum_{i=1}^3\hat{t}_i)^{-1}$.  Following the original formulation of Skorniakov and Ter-Martirosian \cite{skorniakov1956gv}, we make the ansatz $|\Psi_\nu^{(1)}\rangle\propto \hat{G}^{(0)}_\mathrm{3B}(\tilde{E}^{(\nu)}_\mathrm{3B})\hat{B}_1(|\zeta\rangle\otimes|\mathcal{F}_\nu\rangle)$.  The tensor product is defined as $\langle {\bf q_1},{\bf p_1}(|\zeta\rangle\otimes|\mathcal{F}_\nu\rangle)=\zeta(2q_1)\mathcal{F}_\nu(p_1)$ in terms of the Jacobi vectors ${\bf q_1}=({\bf k_2}-{\bf k_3})/2$ and ${\bf p_1}=(2{\bf k_1}-{\bf k_2}-{\bf k_3})/3$.  This yields the integral equation for the amplitude
\begin{widetext}
\begin{equation}
\label{eq:stmmedium}
\mathcal{F}_\nu(p_1)=2g\tau\left(\tilde{E}_\mathrm{3B}^{(\nu)}-\frac{3\hbar^2 p_1^2}{4m}\right)\int \frac{d^3 p'}{(2\pi)^3}\ (1+\rho_{{\bf p_1}}+\rho_{{\bf p_1+p'}})\frac{\zeta\left(\left|2{\bf p_1}+{\bf p'}\right|\right)\zeta\left(\left|2{\bf p'}+{\bf p'}\right|\right)}{\tilde{E}_\mathrm{3B}^{(\nu)}-\frac{\hbar^2}{m}\left(p_1^2+p'^2+{\bf p_1}\cdot {\bf p'}\right)}\mathcal{F}_\nu(p'),
\end{equation} 
\end{widetext}
where $\tau(z)=1/(1-g\langle\zeta|\hat{G}^{(0)}_\mathrm{2B}(z)\hat{B}|\zeta\rangle)$.  At unitarity, non-trivial  solutions of Eq.~(\ref{eq:stmmedium}) for negative energies correspond to the spectrum of Efimov clusters \cite{SM}.  

Solving Eqs.~(\ref{eq:embeddedt}) and (\ref{eq:stmmedium}), we obtain the evolution of two-body and Efimov cluster binding energies over a range of densities shown in Fig.~\ref{fig:energy_scaling}, where scaling laws are apparent.  Over the time range considered, the two-body bound cluster binding energy scales continuously as a density power law $n^{2/3}$.  At early times ($t\ll t_\mathrm{n}$), however, the ground, first, and second-excited Efimov cluster binding energies $\tilde{E}_\mathrm{3B}^{(0)},$ $\tilde{E}_\mathrm{3B}^{(1)}$, and $\tilde{E}_\mathrm{3B}^{(2)}$, respectively, are insensitive to density variations, displaying the intrinsic discrete scaling of Efimov states in vacuum with the van der Waals energy $E_\mathrm{vdW}=\hbar^2/mr_\mathrm{vdW}^2$.  The initial Efimov cluster spectrum is $|\tilde{E}_\mathrm{3B}^{(\nu)}|=e^{-2\pi\nu/s_0}\hbar^2\kappa_*^2/m$, where $\kappa_*$ is the three-body parameter $\kappa_*\approx 0.211/r_\mathrm{vdW}$ \cite{PhysRevLett.108.263001,PhysRevA.90.022106}.

As the unitary Bose gas forms on the Fermi timescale, Efimov clusters become increasingly sensitive to the background buildup of pairing excitations at the scale of the interparticle spacing.  Generally, Efimov clusters must be more bound than the embedded atom-molecule threshold at energy $\tilde{E}_\mathrm{2B}^{(\mathrm{D})}$ relative to the embedded three-body continuum.  In Fig.~\ref{fig:energy_scaling}, we see that Efimov clusters sensitive to these scales become progressively localized as their binding energies are blueshifted.  Consequently, Efimov clusters scale continuously with the $n^{2/3}$ power law over a range of atomic densities.  This behavior persists until a blueshifted Efimov cluster is either absorbed into the embedded atom-molecule scattering continuum or $a_\mathrm{eff}$ approaches its asymptotic limit as the gas equilibrates.  This process is repeated log-periodically for densities separated by powers of $e^{3\pi/s_0}\approx22.7^3$.  Over the density range of experimental interest, the ground state Efimov cluster energy in Fig.~\ref{fig:energy_scaling}, however, remains insensitive to both density variation and evolution at unitarity due to its relative localization.  

The absorption of an Efimov cluster into the embedded atom-molecule scattering continuum is analogous to the behavior of the vacuum Efimov state spectrum for decreasing $a>0$ \cite{d2017few,0034-4885-80-5-056001}, and therefore we expect this dynamical process to be {\it sequential.}  Although only the first three Efimov clusters are shown in Fig.~\ref{fig:energy_scaling}, our results confirm this behavior also for highly-excited Efimov clusters.  Quantitatively, we estimate absorption times for the excited Efimov clusters at a given density
 \begin{equation}\label{eq:fittedmergescal}
 \frac{t^{(\nu)}(n)}{t_\mathrm{n}}=\left(-0.461+(0.093\pm 0.007) r_\mathrm{vdW} k_\mathrm{n} e^{\nu\pi/s_0}\right)^{-1/2},
 \end{equation} 
 where the uncertainty is due to the finite time-step of our many-body simulation \cite{SM}.  To make Eq.~(\ref{eq:fittedmergescal}) well-defined, we restrict the domain of $t^{(\nu)}(n)$ to densities above $n^{(\nu)}r_\mathrm{vdW}^3=(2.12\pm0.45)\times e^{-3\pi\nu/s_0}$, where $t^{(\nu)}(n^{(\nu)})/t_\mathrm{n}\to\infty$.  For densities below $n^{(\nu)}$, our results indicate that the $\nu^\text{th}$ Efimov cluster remains permanently in the bound-state spectrum.  Furthermore, Eq.~(\ref{eq:fittedmergescal}) predicts that increasingly highly-excited Efimov trimers are absorbed exponentially faster, leaving only a finite number of Efimov clusters on the Fermi timescale.  Due also to the minimal amount of quantum depletion occurring during the quench, $a_\mathrm{eff}$ is initially finite as shown in Fig.~\ref{fig:aeff}, and there is a finite set of Efimov clusters before the sequential absorption commences.

{\it Conclusion.}---By systematically applying the cumulant expansion, we have developed a time-dependent picture of the bound cluster composition of the quenched unitary Bose gas.  The size of the dynamically formed unitary two-body clusters is given by the length scale $a_\mathrm{eff}$, which reduces within a few Fermi times to a value proportional to the inter-particle spacing. As this cluster size governs three-body recombination, it gives rise to a universal per-particle loss rate scaling as $n^2a_\mathrm{eff}^4\propto n^{2/3}$, qualitatively matching the scaling behavior observed experimentally \cite{klauss2017observation,eigen2017universal}.  Analyzing this pathway for three-body recombination remains the subject of future studies.  Through time-resolved spectroscopy at unitarity \cite{Bardon722,PhysRevLett.108.145305,PhysRevLett.108.210406}, the evolution of two- and three-body cluster binding energies might be resolved.  The sensitivity of Efimov clusters to the atomic density on Fermi timescales may be experimentally observable as an oscillation chirp of the three-body Tan contact predicted in Ref.~\cite{PhysRevLett.120.100401}.  Predictions related to three-body correlation dynamics on Fermi timescales require an extension of the cumulant theory presented in this Letter or within the ``excitation picture''   \cite{kira2014excitation,kirancomm,KIRA2015185}.  The study of embedded few-body Schr\"odinger equations may also provide insight into the structure of other systems with substantial quantum depletion \cite{donley2001dynamics,PhysRevLett.96.180405}. 

\begin{acknowledgments}
{\it Acknowledgements.}  The authors thank Jose D'Incao, Murray Holland, John Corson, Paul Mestrom, and Thomas Secker for fruitful discussions.  This work is supported by Netherlands Organisation for Scientific Research (NWO) under Grant 680-47-623.
\end{acknowledgments}

\bibliographystyle{apsrev4-1}
\bibliography{references}

\begin{thebibliography}{67}
\bibitem [{phe()}]{phen}%
  \BibitemOpen
  \href@noop {} {}\bibinfo {note} {We note that the result of
  Refs.~\cite{KIRA2015185,kirancomm} account for third and higher-order
  cumulant effects phenomenologically.}\BibitemShut {Stop}%
\end{thebibliography}
\pagebreak
\begin{widetext}
\begin{center}
\textbf{\large Supplemental Materials: ``Dynamical formation of the unitary Bose gas"}
\end{center}
\end{widetext}
\setcounter{equation}{0}
\setcounter{figure}{0}
\setcounter{table}{0}
\setcounter{page}{1}
\makeatletter
\renewcommand{\theequation}{S\arabic{equation}}
\renewcommand{\thefigure}{S\arabic{figure}}

\section{Two-body Calibration}
We calibrate the two free parameters of our separable pairwise interaction $\Lambda$ and $g$ to reproduce the correct s-wave scattering length $a$ and molecular binding energy away from unitarity.  For separable potentials, as used in the main text, the vacuum two-body T-operator $\hat{T}_\mathrm{2B}$ has a simple form \cite{faddeev2013quantum}
\begin{equation}\label{eq:vacuumtoperator}
\hat{T}_\mathrm{2B}(z)=\frac{g|\zeta\rangle\langle\zeta|}{1-g\langle\zeta|\hat{G}_\mathrm{2B}^{(0)}(z)|\zeta\rangle}.
\end{equation}
The s-wave scattering length $a$ is defined as the zero-energy limit vacuum two-body T-matrix, $\lim_{|{\bf k}|\to0} \langle {\bf k,-k}|\hat{T}_\mathrm{2B}\left(\hbar^2k^2/m+i0\right)|{\bf k',-k'}\rangle=4\pi \hbar^2a/m$ where the notation $+i0$ is shorthand for $\lim_{\epsilon\to0^+} i\epsilon$, and the limit is taken on-shell $|{\bf k}|=|{\bf k'}|$.  Evaluating this limit for Eq.~(\ref{eq:vacuumtoperator}), we obtain
\begin{equation}
g=\frac{4\pi\hbar^2a}{m}\left(1-\frac{2a\Lambda}{\pi}\right)^{-1},
\end{equation}
which was given in the main text as the calibration of $g$. 

\begin{figure*}[t!]
\includegraphics[width=17.2cm]{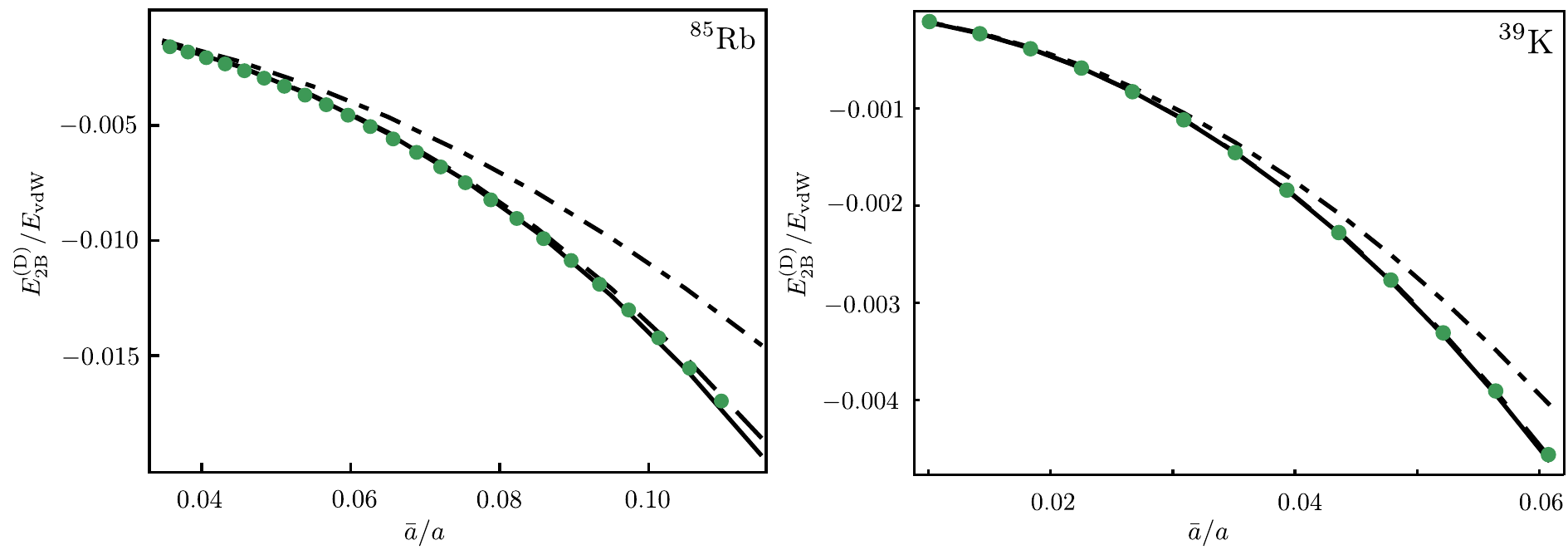}
\caption{\label{fig:molen} Binding energy of the s-wave two-body molecule in vacuum as a function of $a^{-1}$ in units of the mean scattering length $\bar{a}=0.955r_\mathrm{vdW}$ for $^{39}$K (left panel) and $^{85}$Rb (right panel).  We take $r_\mathrm{vdW}=64.61a_0$ for $^{39}$K and $r_\mathrm{vdW}=82.1a_0$ for $^{85}$Rb from Ref.~\cite{RevModPhys.82.1225} where $a_0$ is the Bohr radius.  We compare coupled-channel calculations (green data points) \cite{PhysRevLett.89.180401,thomas} using experimental input from Refs.~\cite{donley2002atom,1367-2630-9-7-223}, the van der Waals correction to the binding energy \cite{PhysRevA.48.546} (dashed), the zero-range limit $\Lambda\to\infty$ of our two-body model (dashed-dotted), and the calibrated result $\Lambda=2/\pi{\bar a}$ within our model (solid).}  
\end{figure*}

To calibrate $\Lambda$, we match finite-range corrections of the molecular binding energy near a broad Feshbach resonance.  To estimate the molecular binding energy within our model, we solve for the location $z=-\hbar^2k_\mathrm{D}^2/m$ of the simple pole in Eq.~(\ref{eq:vacuumtoperator}), which yields the transcendental equation
\begin{equation}\label{eq:trans}
\frac{k_\mathrm{D}}{\Lambda}\arctan\left(\frac{\Lambda}{k_\mathrm{D}}\right)-\frac{\pi}{2a\Lambda}=0.
\end{equation}
Near resonance, $k_\mathrm{D}/\Lambda\ll 1$, and we expand Eq.~(\ref{eq:trans}) to second-order in this small parameter, which may then be solved analytically
\begin{equation}
k_\mathrm{D}=\frac{\pi\Lambda}{4}-\frac{\sqrt{\pi^2a\Lambda^2-8\Lambda\pi}}{4\sqrt{a}}.
\end{equation}    
Equating this correction to the molecular binding energy with the van der Waals correction to the s-wave binding energy $-\hbar^2/m(a-\bar{a})^2$ \cite{PhysRevA.48.546} and expanding in the small parameter $\bar{a}/a$, we obtain $\Lambda=2/\pi{\bar a}$ which was given in the main text as the calibration of $\Lambda$.  In Fig.~\ref{fig:molen}, the prediction for the binding energy within our calibrated two-body model is compared to coupled-channel calculations for the binding energy of the Feshbach molecule for $^{85}$Rb near the resonance at $155$ G \cite{PhysRevLett.89.180401} and $^{39}$K near the resonance at $402$ G \cite{thomas}.  We find generally good agreement near resonance with the coupled-channel calculation using our calibration scheme compared to the zero-range limit $\Lambda\to\infty$ shown in Fig.~\ref{fig:molen}, which justifies our approach.   

\section{Hartree-Fock-Bogoliubov Theory of the Quenched Unitary Bose-condensed gas}\label{SecI}
Following Refs.~\cite{blaizot1986quantum,proukakis1996generalized}, we derive the Hartree-Fock-Bogoliubov equations of motion from the energy functional
\begin{eqnarray}
\langle\hat{H}\rangle&=&\hat{E}[\psi_0,\kappa,\rho]=\sum_{{\bf k}\neq 0}\frac{\hbar^2k^2}{2m}\rho_{\bf k}+2g\sum_{{\bf k}\neq 0}|\zeta({\bf k})|^2|\psi_0|^2\rho_{\bf k}\nonumber\\
&+&\frac{g}{2}\sum_{{\bf k}\neq 0}\zeta(0)\zeta^*(2{\bf k})(\psi_0^*)^2\kappa_{\bf k}+(\text{c.c})\nonumber\\
&+&\frac{g}{2}\sum_{\{{\bf k,k'\}\neq0}}\left[\zeta(2{\bf k})\zeta^*(2{\bf k'})\kappa_{\bf k}\kappa^*_{\bf k'}+2|\zeta({\bf k}-{\bf k'})|^2\rho_{\bf k}\rho_{\bf k'}\right]\nonumber\\
\end{eqnarray}
including up to second-order cumulants, where $\kappa_{ij}=\langle \hat{a}_{{\bf k}_j}\hat{a}_{{\bf k}_i}\rangle_c$ and $\rho_{ij}=\langle \hat{a}^\dagger_{{\bf k}_j}\hat{a}_{{\bf k}_i}\rangle_c$ are the pair and one-body density matrices for excitations $({\bf k}\neq 0)$, respectively.  From functional derivatives of the energy functional, we define the pairing field and Hartree-Fock Hamiltonian
\begin{eqnarray}
\Delta_{ij}\equiv\frac{\delta \hat{E}[\psi_0,\kappa,\rho]}{\delta \kappa^*_{ji}},\\
h_{ij}\equiv\frac{\delta \hat{E}[\psi_0,\kappa,\rho]}{\delta \rho_{ji}},
\end{eqnarray}
from which we define the quasiparticle Hamiltonian
\begin{equation}
\mathcal{H}\equiv\left(
\begin{array}{cc}
h  & \Delta \\
-\Delta^* & -h^*\\
\end{array}
\right),
\end{equation}
and generalized one-body density matrix
\begin{equation}
\mathcal{R}\equiv\left(
\begin{array}{cc}
\rho & \kappa\\
\kappa^* & (\rho+1)\\
\end{array}
\right).
\end{equation}
Now, the second-order cumulant equations of motion can be written simply as
\begin{eqnarray}
i\hbar\dot{\psi}_0&=&\frac{\delta \hat{E}[\psi_0,\kappa,\rho]}{\delta \psi_0^*}\\
i\hbar\dot{\mathcal{R}}&=&\mathcal{H}\mathcal{R}-\mathcal{R}\mathcal{H}^\dagger,
\end{eqnarray}
which results in the equations of motion, Eqs.~(3)--(5), given in the main text.  This second-order cumulant theory is equivalent to the many-body formalism used in Ref.~\cite{PhysRevA.89.021601} to study the quenched unitary Bose-condensed gas as suggested in Ref.~\cite{PhysRevA.90.021602}.  

To simulate a quench experiment for a uniform gas, we begin with an initially pure Bose-Einstein condensate, and then ramp $a\to\infty$ over $5$ $\mu$s, following the approach used in Ref.~\cite{PhysRevA.89.021601}.  We then evolve the gas in the unitary regime until the assumptions underlying our model are violated as $\rho_{\bf k}$ exceeds unity and quantum depletion becomes significant.  This occurs after evolving roughly $t\sim2t_\mathrm{n}$ in the unitary regime.  Provided $nr_\mathrm{vdW}^3\ll1$, we confirm the universal behavior of $\rho_{\bf k}$ observed in Ref.~\cite{PhysRevA.89.021601} as the gas evolves in the unitary regime.  We then take the universal evolution of $\rho_{\bf k}$ as input into the embedded few-body Schr\"odinger equations in our calculation of the bound few-body clusters.  To resolve $\rho_{\bf k}$, which is a function only of the magnitude $|{\bf k}|$ in our translationally invariant system, we use 20000 k-space gridpoints evenly-spaced on the interval $k\in[0,\Lambda]$.  Truncating the grid at $\Lambda$ is justified provided $\rho_{\bf \Lambda}\ll1$, which we find to hold provided $nr_\mathrm{vdW}^3\ll1$.  To go beyond $t\sim 2t_\mathrm{n}$, we must move to a number-conserving approach like the excitation picture \cite{kira2014excitation} or investigate whether introducing higher-order correlations or inelastic losses within our formalism slows the progression of quantum depletion.   

\section{Embedded Two-body problem}\label{sec:two}
The cumulant equation of motion for $\kappa_{\bf k}$ can be written as
\begin{equation}
i\hbar \dot{\kappa}_{\bf k}=2h({\bf k})\kappa_{\bf k}+(1+2\rho_{\bf k})g\zeta(2{\bf k})\sum_{{\bf k}\neq 0}\zeta^*(2{\bf k'})\kappa_{\bf k'}+\Gamma^{0,2}_{\bf k},\label{eq:kappadot}
\end{equation}
where the effect of inhomogeneities is summarized by $\Gamma^{0,2}_{\bf k}$.  Treating density effects as quasi-stationary, we repeatedly solve for the stationary states satisfying
\begin{align}
E^{(\nu)}_\mathrm{2B}\phi_\nu({\bf k})&=2h({\bf k})\phi_\nu({\bf k})&\nonumber\\
&+(1+2\rho_{\bf k})\sum_{{\bf k'}\neq 0}g\zeta(2{\bf k})\zeta^*(2{\bf k'})\phi_\nu({\bf k'})+\Gamma^{0,2}_{\bf k},&\nonumber\\\label{eq:stationary2}
\end{align}
over the course of a many-body simulation.  To study the principle structure of Eq.~(\ref{eq:stationary2}), we ignore $\Gamma^{0,2}_{\bf k}$ which yields effects such as effective decay and secondary energy shifts \cite{kira2011semiconductor,KIRA2015185}.  This results in the real-valued eigenvalue problem given by Eq.~(8) of the main text.  

To solve Eq.~(8), we begin by defining the two-body embedded Green's operator $\hat{G}_\mathrm{2B}(z)\equiv(z-(2\hat{t}+\hat{B}\hat{V}))^{-1}$ where the Bose-enhancement operator $\hat{B}$ is defined as $\langle {\bf k,k'}|\hat{B}=(1+\rho_{\bf k}+\rho_{\bf k'})\langle {\bf k,k'}|$.  The two-body embedded Green's operator satisfies the Lippman-Schwinger equation \cite{taylor2006scattering}
\begin{align}\label{eq:lsg2}
\hat{G}_\mathrm{2B}(z)&=\hat{G}_\mathrm{2B}^{(0)}(z)+\hat{G}_\mathrm{2B}^{(0)}(z)\hat{B}\hat{V}\hat{G}_\mathrm{2B}(z),&\nonumber\\
&=\hat{G}_\mathrm{2B}^{(0)}(z)+\hat{G}_\mathrm{2B}(z)\hat{B}\hat{V}\hat{G}_\mathrm{2B}^{(0)}(z).&
\end{align}
We then define the embedded two-body T-operator 
\begin{equation} 
\hat{\mathcal{T}}_\mathrm{2B}(z)\equiv\hat{B}\hat{V}+\hat{B}\hat{V}\hat{G}_\mathrm{2B}\hat{B}\hat{V},
\end{equation}
which has the same properties as $\hat{G}_\mathrm{2B}(z)$ as an analytic function of $z$.  This is analogous to the vacuum definition of the two-body T-operator \cite{taylor2006scattering}. We obtain the identities $\hat{G}^{(0)}_\mathrm{2B}\hat{\mathcal{T}}_\mathrm{2B}(z)=\hat{G}_\mathrm{2B}(z)\hat{B}\hat{V}$ and $\hat{\mathcal{T}}_\mathrm{2B}(z)\hat{G}^{(0)}_\mathrm{2B}(z)=\hat{B}\hat{V}\hat{G}_\mathrm{2B}(z)$ straightforwardly.  From these identities, we obtain the Lippman-Schwinger equation 
\begin{equation}\label{eq:lst2}
\hat{\mathcal{T}}_\mathrm{2B}(z)=\hat{B}\hat{V}+\hat{B}\hat{V}\hat{G}^{(0)}_\mathrm{2B}(z)\hat{\mathcal{T}}_\mathrm{2B}(z),
\end{equation}
given in the main text.  Our embedded two-body T-operator is related to the many-body T-operator of Ref.~\cite{stoof1996theory} via $\hat{B}\hat{T}_\mathrm{MB}(z)=\hat{\mathcal{T}}_\mathrm{2B}(z)$, and therefore also has the same analytic properties as $\hat{G}_\mathrm{2B}(z)$.  The many-body T-operator is related to the vacuum two-body T-operator as $\hat{T}_\mathrm{MB}(z)=\hat{T}_\mathrm{2B}(z)+\hat{T}_\mathrm{2B}(z)\hat{G}_\mathrm{2B}^{(0)}(z)(\hat{B}-1)\hat{T}_\mathrm{MB}(z)$ \cite{stoof1996theory}. 

\begin{figure*}[t!]
\includegraphics[width=17.2cm]{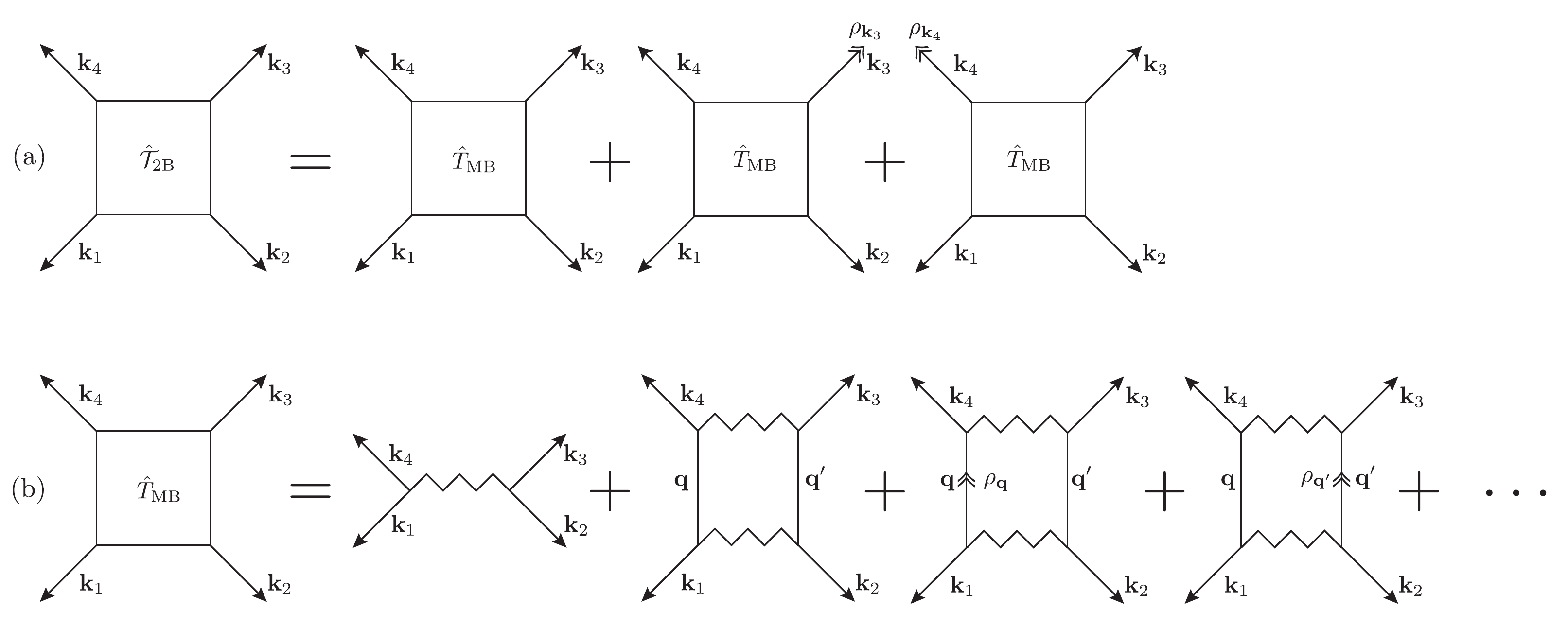}
\caption{\label{fig:feyn}  (a) An illustration of the decomposition $\hat{\mathcal{T}}_\mathrm{2B}=\hat{B}\hat{T}_\mathrm{MB}$, including Bose-stimulation of collision outputs indicated by double arrows.  (b) Diagrammatic representation of the Born series for $\hat{T}_\mathrm{MB}$ in terms of Feynman diagrams.  The jagged lines indicate a pairwise interaction.}
\end{figure*}

The Born series for $\hat{\mathcal{T}}_\mathrm{2B}(z)$ can be interpreted graphically as Feynman diagrams shown in Fig.~\ref{fig:feyn}, although its convergence is not guaranteed \cite{taylor2006scattering}.  To obtain the general closed-form expression for $\hat{\mathcal{T}}_\mathrm{2B}(z)$, we begin by writing Eq.~(\ref{eq:lst2}) for a separable pairwise potential $\hat{V}=g|\zeta\rangle\langle\zeta|$, giving
\begin{equation}\label{eq:sept}
\hat{\mathcal{T}}_\mathrm{2B}(z)=g\hat{B}|\zeta\rangle\langle\zeta|+g\hat{B}|\zeta\rangle\langle\zeta|\hat{G}_\mathrm{2B}^{(0)}(z)\hat{\mathcal{T}}_\mathrm{2B}(z).
\end{equation}
Applying $\langle \zeta|G_\mathrm{2B}^{(0)}(z)$ to the left hand side of Eq.~(\ref{eq:sept}), we obtain 
\begin{equation}\label{eq:lhs}
\langle \zeta|\hat{G}_\mathrm{2B}^{(0)}(z)\hat{\mathcal{T}}_\mathrm{2B}(z)=\frac{g\langle\zeta|\hat{G}_\mathrm{2B}^{(0)}(z)\hat{B}|\zeta\rangle}{1-g\langle\zeta|\hat{G}_\mathrm{2B}^{(0)}(z)\hat{B}|\zeta\rangle}\langle\zeta|.
\end{equation}
Inserting Eq.~(\ref{eq:lhs}) into Eq.~(\ref{eq:sept}), yields Eq.~(10) of the main text
\begin{equation} \label{eq:2bt}
\hat{\mathcal{T}}_\mathrm{2B}(z)=\hat{B}\frac{g|\zeta\rangle\langle\zeta|}{1-g\langle\zeta|\hat{G}_\mathrm{2B}^{(0)}(z)\hat{B}|\zeta\rangle}.
\end{equation}
We obtain the embedded two-body cluster energies by locating the simple pole in Eq~.\ref{eq:2bt}.  The integrals in the denominator of Eq.~\ref{eq:2bt} are evaluated by Gaussian quadrature \cite{press1989numerical} and interpolation of $\rho_{\bf k}$ from the dense simulation grid onto a grid of $2000$ abscissas distributed on the interval $k\in[0,\Lambda]$.  

\section{Embedded Three-body problem}
The cumulant equation of motion for $\tau^{0,3}_{\bf k,k'}$ can be written as
\begin{widetext}
\begin{equation}
i\hbar \dot{\tau}^{0,3}_{\bf k,k'}=(1+\hat{P}_++\hat{P}_-)\left(h({\bf k}) \tau^{0,3}_{\bf k,k'}+(1+\rho_{\bf k'}+\rho_{{\bf k}+{\bf k'}})\sum_{{\bf k''}\neq 0}g\zeta(2{\bf k'}+{\bf k})\zeta^*(2{\bf k''}+{\bf k}) \tau^{0,3}_{\bf k,k''}+\Gamma^{0,3}_{\bf k,k'}\right),
\end{equation}
\end{widetext}
where the effect of inhomogeneities is summarized by $\Gamma^{0,3}_{\bf k,k'}$.  We treat density effects as quasi-stationary and ignore $\Gamma^{0,3}_{\bf k,k'}$, which gives Eq.~(9) of the main text.  Equation~(9) is then repeatedly solved for the stationary states $\Psi_\nu({\bf k,k'})$ over the course of a many-body simulation.

To solve Eq.~(9), we follow the original formulation of Skorniakov and Ter-Martirosian \cite{skorniakov1956gv} to derive Eq.~(12) of the main text.  We begin from the Faddeev equation \cite{faddeev2013quantum} for a three-body bound state with effective pairwise interaction $\hat{V}_\mathrm{eff}$
\begin{equation}\label{eq:stm1}
 |\Psi^{(1)}_\nu\rangle=\hat{G}_\mathrm{3B}^{(0)}(\tilde{E}_\mathrm{3B}^{(\nu)})\hat{\mathcal{T}}_{23}(\tilde{E}_\mathrm{3B}^{(\nu)})(\hat{P}_++\hat{P}_-)|\Psi^{(1)}\rangle,
 \end{equation}
 where $\hat{\mathcal{T}}_{23}(z)$ is defined in the main text.  First, we rewrite Eq.~(\ref{eq:stm1}) in momentum space using the Jacobi coordinates
 \begin{widetext}
 \begin{equation}
 \Psi^{(1)}_\nu({\bf q_1},{\bf p_1})=G_\mathrm{3B}^{(0)}(q_1,p_1,\tilde{E}_\mathrm{3B}^{(\nu)})\int\frac{d^3q'}{(2\pi)^3}\int\frac{d^3p'}{(2\pi)^3}\langle {\bf q_1},{\bf p_1}|\hat{\mathcal{T}}_{23}(\tilde{E}_\mathrm{3B}^{(\nu)})|{\bf q'},{\bf p'}\rangle\langle {\bf q'},{\bf p'}|\hat{P}_++\hat{P}_-| \Psi^{(1)}_\nu\rangle.
 \end{equation}
 \end{widetext}
 For a separable potential, this integral equation may be further simplified by using the result of Eq.~(\ref{eq:2bt}) 
 \begin{widetext}
 \begin{align}
 \langle {\bf q_1},{\bf p_1}|\hat{\mathcal{T}}_{23}(\tilde{E}_\mathrm{3B}^{(\nu)})|{\bf q'},{\bf p'}\rangle&=g\delta^{(3)}({\bf p_1}-{\bf p'})(1+\rho_{{\bf q_1}-{\bf p_1}/2}+\rho_{{\bf q_1}+{\bf p_1}/2})\zeta(2{\bf q_1})\zeta^*(2{\bf q'})\tau\left( \tilde{E}_\mathrm{3B}^{(\nu)}-\frac{3\hbar^2p_1^2}{4m}\right),&\\
 \Psi^{(1)}_\nu({\bf q_1},{\bf p_1})&=g(1+\rho_{{\bf q_1}-{\bf p_1}/2}+\rho_{{\bf q_1}+{\bf p_1}/2})G_\mathrm{3B}^{(0)}(q_1,p_1,\tilde{E}_\mathrm{3B}^{(\nu)})\tau\left( \tilde{E}_\mathrm{3B}^{(\nu)}-\frac{3\hbar^2p_1^2}{4m}\right)\zeta(2{\bf q_1})&\nonumber\\
 &\times\int \frac{d^3p'}{(2\pi)^3}\zeta^*(2{\bf p'}+{\bf p_1})\Psi^{(1)}_\nu({\bf q_1},{\bf p_1})).&
 \end{align}
 \end{widetext}
 Now, we make the ansatz
 \begin{equation}\label{eq:stm2}
 |\Psi^{(1)}_\nu\rangle=N\hat{G}^{(0)}_\mathrm{3B}(\tilde{E}_\mathrm{3B}^{(\nu)})\hat{B}_1(|\zeta\rangle\otimes|\mathcal{F}_\nu\rangle),
 \end{equation}
 where $N$ is the normalization constant, and $\langle {\bf q_1},{\bf p_1}|\hat{B}_1=\langle{\bf q_1},{\bf p_1}|(1+\rho_{{\bf q_1}-{\bf p_1}/2}+\rho_{{\bf q_1}+{\bf p_1}/2})$ is the Bose-enhancement operator using spectator notation in terms of Jacobi coordinates.  Inserting this ansatz into Eq.~(\ref{eq:stm1}), and for s-wave pairwise interactions, we obtain the amplitude
 \begin{widetext}
\begin{equation}
\mathcal{F}_\nu(p_1)=2g\tau\left(\tilde{E}_\mathrm{3B}^{(\nu)}-\frac{3\hbar^2 p_1^2}{4m}\right)\int \frac{d^3 p'}{(2\pi)^3}\ (1+\rho_{{\bf p_1}}+\rho_{{\bf p_1+p'}})\frac{\zeta\left(\left|2{\bf p_1}+{\bf p'}\right|\right)\zeta\left(\left|2{\bf p'}+{\bf p'}\right|\right)}{\tilde{E}_\mathrm{3B}^{(\nu)}-\frac{\hbar^2}{m}\left(p_1^2+p'^2+{\bf p_1}\cdot {\bf p'}\right)}\mathcal{F}_\nu(p'),
\end{equation} 
\end{widetext}
 which is Eq.~(12) of the main text.   

\section{Calculation of Absorption Times}
To calculate Eq.~(13) for the scaling laws obeyed by the absorption times $t^{(\nu)}$, we numerically estimate $t^{(\nu)}$ over a range of densities, observing that the $\nu^\text{th}$ excited Efimov cluster is absorbed when $a_\mathrm{eff}=a^{(\nu)}$, where $a^{(\nu)}$ is approximately density-independent.  Estimating $a^{(2)}=(165\pm12)r_\mathrm{vdW}$, due to uncertainty in $t^{(\nu)}$ from the finite time-step of the many-body simulation, we establish the general scaling $a^{(\nu+1)}\approx e^{\pi/s_0}a^{(\nu)}$ for excited Efimov clusters.  Using the universal result in Eq.~(11), we then extend this result to characterize $t^{(\nu)}$ at arbitrary densities, resulting in Eq.~(13).    

\begin{acknowledgments}
{\it Acknowledgements.}  The authors thank Thomas Secker for providing us with molecular binding energies for $^{39}$K from a coupled-channels calculation \cite{thomas}.  
\end{acknowledgments}


\end{document}